\documentclass[journal]{IEEEtran} 
\usepackage{graphicx} 
\usepackage{xcolor}
\usepackage{amssymb}
\usepackage{multirow}
\usepackage[hidelinks]{hyperref}
\begin{document}
	\title{Dimension reduction methods, persistent homology and machine learning for EEG signal analysis of Interictal Epileptic Discharges}

	\author{Annika Stiehl, Stefan Geißelsöder, Nicole Ille, Fabienne Anselstetter, Harald Bornfleth, Christian Uhl%
		\thanks{A. Stiehl, S. Geißelsöder and C. Uhl are with the University of Applied Sciences Ansbach, Residenzstraße 8, Germany,\\
			E-Mail: \href{mailto:a.stiehl18120@hs-ansbach.de}{a.stiehl18120@hs-ansbach.de.}
		}
		\thanks{N. Ille, F. Anselstetter and H. Bornfleth are with BESA GmbH, Freihamer Straße 18, Gräfelfing, Germany.}
	}
	
	\markboth{Proceedings of the WORKSHOP BIOSIGNAL 2022, August 24th\,-\,26th, 2022, Dresden, Germany}%
	{Kurths \MakeLowercase{\textit{et al.}}: Please copy your title here}
	
	\maketitle

	\begin{abstract}
		Recognizing specific events in medical data requires trained personnel. To aid the classification, machine learning algorithms can be applied. In this context, medical records are usually high-dimensional, although a lower dimension can also reflect the dynamics of the signal. In this study, electroencephalogram data with Interictal Epileptic Discharges (IEDs) are investigated. First, the dimensions are reduced using Dynamical Component Analysis (DyCA) and Principal Component Analysis (PCA), respectively. The reduced data are examined using topological data analysis (TDA), specifically using a persistent homology algorithm. The persistent homology results are used for targeted feature generation. The features are used to train and evaluate a Support Vector Machine (SVM) to distinguish IEDs from background activities. 
	\end{abstract}
	
	
	\IEEEpeerreviewmaketitle

	\section{Introduction}
	The evaluation of medical data is increasingly automated. However, it is important that the evaluation is comprehensible and the path of the decision can be interpreted. Feature engineering facilitates accessibility to machine learning models and is one of the most frequently used data preprocessing methods \cite{Galli2021}. One possibility of feature engineering will be presented in this paper.
	
	A combination of dimensionality reduction and persistent homology to generate interpretable features is investigated. The data examined are electroencephalography (EEG) data, which contain Interictal Epileptic Discharges (IED). Two dimension reduction methods are used; one is Principal Component Analysis (PCA) \cite{Pearson1901} and the other is the Dynamical Component Analysis (DyCA) \cite{Seifert2018}. The two methods differ in that DyCA leads to a generalized eigenvalue problem focusing on the dynamic components, whereas PCA is based on a statistical model assumption. 
	%
	\section{Data}
	The data for this study is taken from the Temple University Hospital EEG Corpus \cite{EEGCorpus}. It includes EEG recordings of 24 to 36 channels from epileptic patients of which IEDs are investigated. For this purpose, five different events are available in the processed and filtered data set. Two of the five events can be assigned to IEDs and form the positive class. The remaining event types are background activities and form the negative class for this study. The raw data were resampled to 128 Hz, bandpass filtered and remontaged to different virtual EEG montages using the commercially available software package BESA Research (BESA GmbH, Gräfelfing, Germany). In this study the bipolar (28 channels), the average (27 channels) and the Cz reference montage (27 channels) were examined. 
	\section{Methods}
	\subsection{Dynamical Component Analysis}
	The Dynamical Component Analysis (DyCA) is a dimension reduction method based on modeling the simulated or measured signal by a series of ordinary differential equations (ODEs) \cite{Seifert2018}. The principal aim of the algorithm is to split a signal $q(t) \in \mathbb{R}^N$, where $N$ describes the number of dimensions, into a deterministic part, $\sum x_i(t) w_i$, and an independent noise component, $\sum \xi_j(t) \psi_j$: 
	\begin{equation}\label{eq:1}
		q(t) = \sum_{i=1}^n x_i(t)w_i+\sum_{j=1}^p \xi_j(t)\psi_j
	\end{equation} 
	assuming that $n+p\le N$ and $w_i$, $\psi_j$ being linearly independent. The dynamics of the deterministic amplitudes $x_i(t)$ of the signal $q(t)$ is assumed to be driven by a system of $m$ linear and $n-m$ nonlinear differential equations, with $m \geq n-m$. Defining correlation matrices a generalized eigenvalue problem arises. Projecting the high-dimensional signal onto its generalized eigenvectors a low-dimensional signal representation is obtained.
	\subsection{Persistent Homology}
	Persistent homology is one of the standard tools from statistical topological data analysis (TDA) \cite{Nastansky2019}. Its goal is the analysis of data represented by a point cloud in $N$ dimensions. Thereby persistent homology provides a multiscale description of the topology of the point cloud. To calculate the topological features of the given data, the point cloud must be represented as a simplicial complex, in most implementations the Vietoris–Rips complex is used \cite{Bubenik2017}. Various visualization options for the analyses are provided, like the persistence barcodes or persistence landscapes. The persistence landscape (PL) is a sequence of piecewise linear functions
	$\lambda_k  \in \mathbb{R}^+$ with $k \in \mathbb{N}$ from the birth-death pairs of the calculated homology, where $\lambda_k$ is the $k$-th largest value of the birth-death pairs \cite{Bubenik2017}. 
	\section{Application and Results}
	\begin{figure}[!t]
		\centering
		\includegraphics[width=\linewidth]{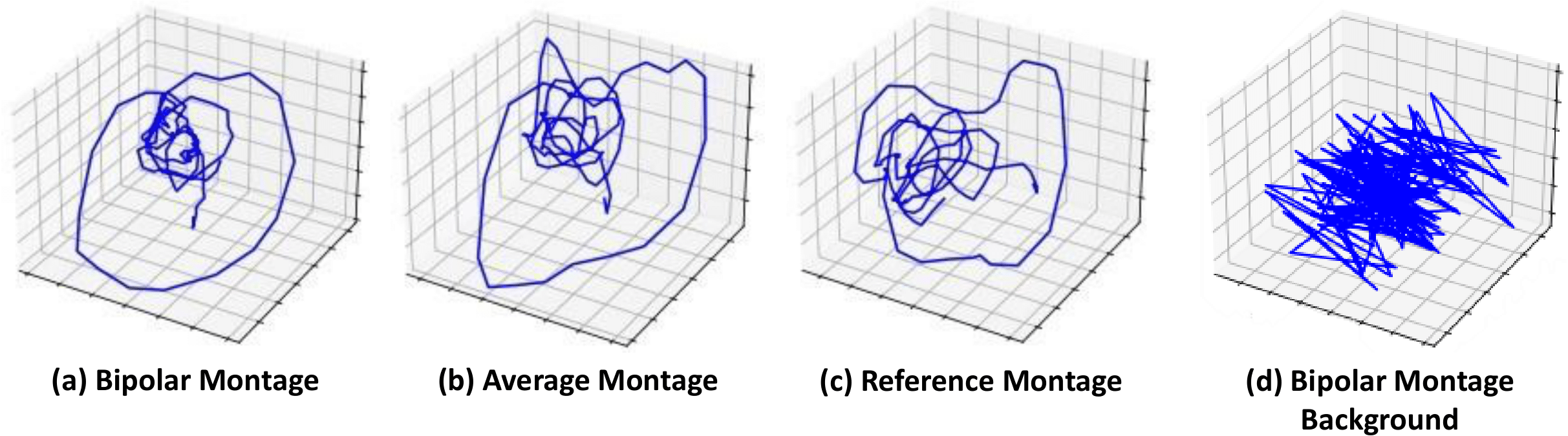}
		\caption{DyCA 3D-trajectories computed from (a) a bipolar montage with 28 Channels, (b) a average montage with 27 channels and (c) a reference montage with 27 channels using the same EEG recording. In each case the same one second section with a IED was selected. (d) shows a one second section of background data in the bipolar montage.}
		\label{fig:montages}
	\end{figure}
	Python is used to perform the experiments with an implementation of DyCA, the ripser, persim and scikit-learn libraries. The first part of the study deals with the analysis of different ways of montage of an EEG signal: the bipolar, average and reference montage. For this purpose, the $N$ dimensional data is read and transformed into a $n$ dimensional signal using the dimensionality reduction methods DyCA and the commonly known PCA \cite{Pearson1901}. For the reduction we choose $n=3$, with knowledge from \cite{Seifert2018a}, where epileptic absences represent a Shilnikov attractor with two linear and one non-linear differential equations. A part of the results can be seen in the Figure \ref{fig:montages}: The dimensionality reduction method DyCA is used and a one second section of the EEG signal during an IED (Figure \ref{fig:montages}~a-c) and some background data (Figure \ref{fig:montages}~(d)) is examined.
	\begin{figure}[!b]
		\centering
		\includegraphics[width=\linewidth]{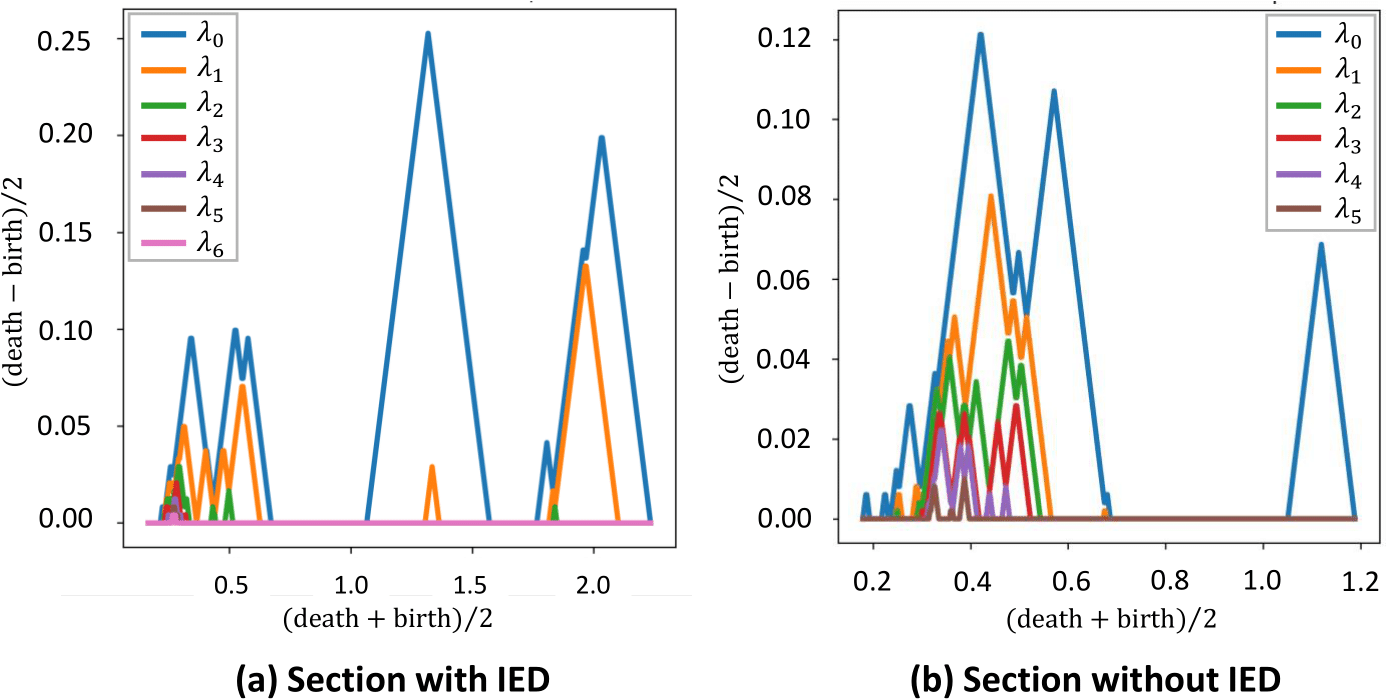}
		\caption{Persistence Landscape of DyCA 3D-trajectories. For both data sections the bipolar montage is used. In (a) a one second section with a IED and in (b) a one second section without IED was chosen.}
		\label{fig:PL}
	\end{figure}
	The figure shows that significant patterns can be visually detected in the case of IEDs. The background data trajectory (Figure \ref{fig:montages}~(d)) can be clearly distinguished visually from the trajectories with an IED. But the different montages of the EEG signals have only minor influence; the significant pattern can be detected in all here used montages. Similar results are obtained when PCA is used instead of DyCA. With the results, further investigations are made, regarding an analysis of the trajectories within a data-driven analysis framework.
	
	In the next step, persistent homology is applied to the three-dimensional data. The results are the birth-death pairs and the representation as PL. Two examples are shown in Figure \ref{fig:PL}, where (a) shows the calculated PL of a one second section with an IED and (b) without an IED. Comparing the two charts, it can be seen that in (a) there are higher values for both the ordinate and the abscissa, suggesting that larger circular structures are found in the associated trajectory (Figure \ref{fig:montages}~(a)) than Figure \ref{fig:PL}~(b) with associated trajectory Figure \ref{fig:montages}~(d).
	
	To gain further insight if these findings are suitable for machine learning applications features are generated using the results from the persistence barcodes and the PL; both statistical features and polynomial features are used. These are calculated for dimension 0 (connections between two components) and for dimension 1 (loop in the data) of the persistent homology. In total there are 40 features each dataset, used to train a Support Vector Classifier (SVC). For this, 1100 one second segments are utilized and 15\% of these are taken as independent evaluation data. The SVC is optimized with a GridSearch and with the help of cross-validation. This results in a training accuracy score of 76 \% and an evaluations accuracy score of 85 \%. The results suggest that using persistent homology, the trajectories can be interpreted and assigned to the associated classes using an SVC.

	\section{Conclusion}
	In these studies, we show that significant patterns are evident in the trajectories after dimension reduction (DyCA and PCA) in the case of an IED. However, the different montages have small influence on the three-dimensional trajectory. Moreover, persistent homology can be applied to the trajectories. With the help of this, features can be generated and applied to a Support Vector Classifier. To get a better understanding of which features of the persistent homology are most relevant, in future studies, a feature selection will be performed before training. 
	
	\section*{Acknowledgment}
	This work was supported by the German Federal Ministry of Education and Research (BMBF, Funding number: 05M20WBA).

	\ifCLASSOPTIONcaptionsoff
	\newpage
	\fi

	
	
	%
	\bibliographystyle{IEEEtran}
	\bibliography{IEEEabrv, biblio.bib}

\begin{thebibliography}{1}
\providecommand{\url}[1]{#1}
\csname url@samestyle\endcsname
\providecommand{\newblock}{\relax}
\providecommand{\bibinfo}[2]{#2}
\providecommand{\BIBentrySTDinterwordspacing}{\spaceskip=0pt\relax}
\providecommand{\BIBentryALTinterwordstretchfactor}{4}
\providecommand{\BIBentryALTinterwordspacing}{\spaceskip=\fontdimen2\font plus
\BIBentryALTinterwordstretchfactor\fontdimen3\font minus
  \fontdimen4\font\relax}
\providecommand{\BIBforeignlanguage}[2]{{%
\expandafter\ifx\csname l@#1\endcsname\relax
\typeout{** WARNING: IEEEtran.bst: No hyphenation pattern has been}%
\typeout{** loaded for the language `#1'. Using the pattern for}%
\typeout{** the default language instead.}%
\else
\language=\csname l@#1\endcsname
\fi
#2}}
\providecommand{\BIBdecl}{\relax}
\BIBdecl

\bibitem{Galli2021}
S.~Galli, ``Feature-engine: A python package for feature engineering for
  machine learning,'' \emph{Journal of Open Source Software}, vol.~6, no.~65,
  p. 3642, sep 2021.

\bibitem{Pearson1901}
K.~Pearson, ``On lines and planes of closest fit to systems of points in
  space,'' \emph{The London, Edinburgh, and Dublin Philosophical Magazine and
  Journal of Science}, vol.~2, no.~11, pp. 559--572, nov 1901.

\bibitem{Seifert2018}
B.~Seifert, K.~Korn, S.~Hartmann, and C.~Uhl, ``Dynamical component analysis
  ({DYCA}): Dimensionality reduction for high-dimensional deterministic
  time-series,'' in \emph{2018 {IEEE} 28th International Workshop on Machine
  Learning for Signal Processing ({MLSP})}.\hskip 1em plus 0.5em minus
  0.4em\relax {IEEE}, sep 2018, pp. 1--6.

\bibitem{EEGCorpus}
\BIBentryALTinterwordspacing
J.~Picone, \emph{Electroencephalography (EEG) Resources}, Temple University
  Hospital. [Online]. Available:
  \url{https://isip.piconepress.com/projects/tuh\_eeg/index.shtml}
\BIBentrySTDinterwordspacing

\bibitem{Nastansky2019}
A.~Nastansky, \emph{\BIBforeignlanguage{de}{Topologische
  {D}atenanalyse}}.\hskip 1em plus 0.5em minus 0.4em\relax Universität
  Potsdam, 2019.

\bibitem{Bubenik2017}
P.~Bubenik and P.~D{\l}otko, ``A persistence landscapes toolbox for topological
  statistics,'' \emph{Journal of Symbolic Computation}, vol.~78, pp. 91--114,
  jan 2017.

\bibitem{Seifert2018a}
B.~Seifert, D.~Adamski, and C.~Uhl, ``Analytic quantification of {S}hilnikov
  chaos in epileptic {EEG} data,'' \emph{Frontiers in Applied Mathematics and
  Statistics}, vol.~4, nov 2018.

\end{thebibliography}
	%

	
\end{document}